\documentclass[%
 reprint,
 superscriptaddress,
 amsmath,amssymb,
 aip,
floatfix,
]{revtex4-1}

\usepackage{graphicx}
\usepackage{dcolumn}
\usepackage{bm}
\usepackage{siunitx}

\begin{document}

\title{Thermal radiation dominated heat transfer in nanomechanical silicon nitride drum resonators}

\author{Markus Piller}
\affiliation{%
 Institute of Sensor and Actuator Systems, TU Wien, Gusshausstrasse 27-29, 1040 Vienna.
}
\author{Pedram Sadeghi}
\affiliation{%
 Institute of Sensor and Actuator Systems, TU Wien, Gusshausstrasse 27-29, 1040 Vienna.
}
\author{Robert G. West}
\affiliation{%
 Institute of Sensor and Actuator Systems, TU Wien, Gusshausstrasse 27-29, 1040 Vienna.
}
\author{Niklas Luhmann}
\affiliation{%
 Institute of Sensor and Actuator Systems, TU Wien, Gusshausstrasse 27-29, 1040 Vienna.
}
\author{Paolo Martini}
\affiliation{%
 Institute of Sensor and Actuator Systems, TU Wien, Gusshausstrasse 27-29, 1040 Vienna.
}
\author{Ole Hansen}
\affiliation{DTU Nanolab, Technical University of Denmark, Kongens Lyngby DK-2800, Denmark}
\author{Silvan Schmid}%
 \email{silvan.schmid@tuwien.ac.at}
\affiliation{%
 Institute of Sensor and Actuator Systems, TU Wien, Gusshausstrasse 27-29, 1040 Vienna.
}

\date{\today}

\begin{abstract}
Nanomechanical silicon nitride (SiN) drum resonators are currently employed in various fields of applications that arise from their unprecedented frequency response to physical quantities. In the present study, we investigate the thermal transport in nanomechanical SiN drum resonators by analytical modelling, computational simulations, and experiments for a better understanding of the underlying heat transfer mechanism causing the thermal frequency response. Our analysis indicates that radiative heat loss is a non-negligible heat transfer mechanism in nanomechanical SiN resonators limiting their thermal responsivity and response time. This finding is important for optimal resonator designs for thermal sensing applications as well as cavity optomechanics. 
\end{abstract}

\maketitle

Since their emergence, nanomechanical resonators have shown distinct advantages in various fields of application due to their high amplitude and frequency response to external physical quantities.\cite{Schmid2016}
In order to achieve optimal performance, material properties are essential, and silicon nitride (SiN) has proven to be well suited for nanomechanical resonators. The large intrinsic stress results in unprecedented high quality factors based on so-called "damping dilution", which has been observed in silicon nitride strings\cite{Verbridge2006,Schmid2011} as well as drums.\cite{Zwickl2008,Wilson2009,Yu2012} The combination of high quality factors and excellent optical properties has made nanomechanical SiN drums interesting devices for cavity optomechanics. \cite{thompson2008strong,Wilson2009} Among other things, they have been used for fundamental research, \cite{Purdy2013} and as transducers between optical and radio wave \cite{Bagci2014} or microwave \cite{Andrews2014} signals. Recent developments of optimized trampoline \cite{pluchar2020cavityfree,PhysRevLett.116.147202,PhysRevX.6.021001} and phononic crystal designs \cite{Tsaturyan2016a,Ghadimi764,Reetz2019} are pushing the quality factors of SiN resonators into the realm of room temperature quantum optomechanical experiments. 

From  cavity optomechanics experiments it is well known that the local heating of the laser causes a frequency detuning of SiN drums.\cite{Friedrich2011,Jockel2011} Such photothermal detuning has been extensively used for sensing applications, such as infrared absorption spectroscopy,\cite{Yamada2013,Biswas2014,Andersen2016,Kurek2017} nanoparticle analysis,\cite{Schmid2014d,Larsen2013ACSNano} single-molecule detection,\cite{Chien2018} and, recently, electromagnetic radiation detection.\cite{Piller2019,Zhang2019} Generally, it has been assumed that heat transfer is dominated by conduction, as it was concluded for nanomechanical torsional paddle resonators.\cite{Zhang2013} Recently, evidence for significant radiative heat transfer in large SiN drums has been presented.\cite{Zhang2020} Despite the proliferation of nanomechanical SiN resonators, the underlying heat transfer mechanisms, which cause the thermal frequency response and ultimately determine the performance limit, has not been studied in detail. 

In this work, we investigate the heat transfer in nanomechanical SiN drum resonators by means of computational simulations and experiments to gain a better understanding of the dominating mechanism. We assume the situation of an experiment under vacuum in which heat convection is negligible, and heat transfer happens solely by radiation and conduction. Our study is conducted by local heating of SiN drums with a laser and analysing the resulting frequency and time response. We show that the frequency response as well as the response time are dominated by radiative heat transfer, which is a function of the lateral size of the drums. This is an important finding to be considered for the optimal design of thermal sensors as well as cavity optomechanics experiments.

The relative frequency shift $\delta f$ for an even temperature change $\Delta T = T-T_0$ from an initial temperature $T_0$ of a resonator under tensile stress $\sigma$ with a resonance frequency $f(T)$, such as a drum or a string, is given by\cite{Larsen2011,Schmid2016}  
\begin{equation}\label{eq:df}
    \delta f = \frac{f(T)}{f(T_0)} =\sqrt{1-\frac{\alpha E (T-T_0)}{\sigma}},
\end{equation}
with the thermal expansion coefficient $\alpha$ and Young's modulus $E$.
This results in the relative temperature responsivity (relative frequency shift per change in temperature) of
\begin{equation}\label{eq:dR}
    \frac{1}{f(T_0)}\frac{\partial f(T)}{\partial T} \bigg\rvert_{T=T_0} = -\frac{\alpha E}{2 \sigma}.
\end{equation}
From (\ref{eq:dR}) it is obvious that the observed temperature induced frequency detuning is enhanced for resonators with a low tensile stress $\sigma$.
Therefore, we performed our study with nanomechanical SiN drum resonators made of low-stress silicon nitride. The drums with a thickness of $h=$~\SI{50}{\nm} are supported by a silicon frame with a thickness of \SI{380}{\um}. We present results from square drums of different sizes $L \times L$ with $L$ = \SIlist{0.5;1.0;2.5;4.0}{\mm}. We used a silicon wafer with silicon-rich SiN grown by low-pressure chemical vapor deposition. The square drum shapes were defined on the backside of the wafer by a standard photolithography process and etched by reactive ion etching. To finally release the drum structures, the silicon wafer was etched through from the backside with potassium hydroxide.
In order to vary the thermal conductivity for comparative measurements, a \SI{50}{\nm} thick aluminum layer was deposited on one side of some SiN drums by means of physical vapor deposition process. The tensile stress of the drums was calculated from the fundamental mode frequency for a mass density of \SI{3000}{\kg\per\meter^3} and \SI{2700}{\kg\per\meter^3} for SiN and Al, respectively.\cite{Schmid2016} 

\begin{figure}
\includegraphics{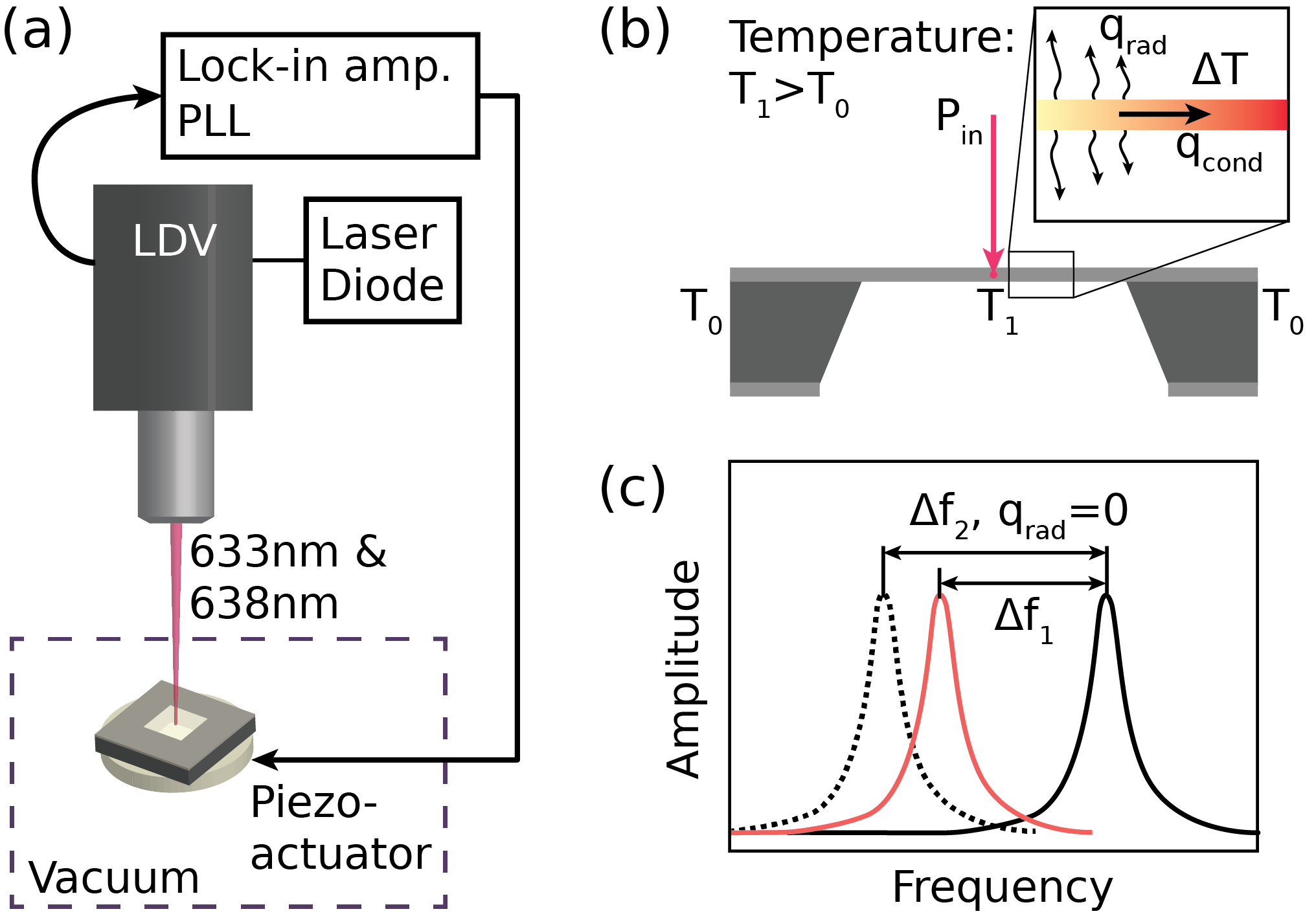}
\caption{\label{fig:fig1} Schematic depiction of (a) the measurement setup, (b) the thermal transport mechanisms for a SiN drum resonator with radiative heat loss from the surface combined with thermal conduction evoked by a temperature gradient from heating the center of the SiN drum by an input laser power $P_{in}$. The surroundings and the Si supporting frame are at room temperature $T_0$ = \SI{300}{\K}. (c) The resonator detuning is comparably smaller considering a radiative heat loss.}
\end{figure}

The experimental setup, schematically depicted in Fig.~\ref{fig:fig1}(a), shows the silicon nitride drum resonator on a piezo actuator and a laser-Doppler vibrometer (LDV) (MSA-500 from Polytec GmbH) to readout the vibrational motion. The signal from the vibrometer is fed into a lock-in amplifier (HF2LI from Zurich Instruments) with an integrated phase-locked loop to control the piezo actuator and to drive the drum at its resonance frequency. A power controllable laser diode (LPS-635-FC from Thorlabs GmbH), with a center wavelength $\lambda = 638$ nm, was attached to the LDV unit to photothermally heat the drums. The exact laser power values were recorded using a silicon photodiode (S120C from Thorlabs GmbH). Simulations shown in this work are performed with COMSOL Multiphysics Version 5.5. All measurements were conducted in a high vacuum at a pressure below \SI{1d-5}{\milli\bar}.

Fig.~\ref{fig:fig1}(b) schematically depicts the heat flux in a drum resonator when locally heated in its center. The incident laser with power $P_{\mathrm{in}}$ is absorbed by the drum, producing a heating power $P_{\mathrm{abs}}= A_{\lambda} P_{\mathrm{in}}$ for a wavelength specific absorbance $A_{\lambda}$. 
According to Fourier's law, the resulting temperature gradient across the drum causes a conductive heat flux $q_{\mathrm{cond}}= -\kappa dT/dx$ from the drum center towards the frame, for a specific thermal conductivity $\kappa$.
The radiative heat transfer from the drum surface is given by the Stefan-Boltzmann law $q_{\mathrm{rad}}=\varepsilon \sigma_{\mathrm{SB}} (T^4 - T_0^4)$, for the special case of having a large surrounding at temperature $T_0$ and the assumption of a gray surface with an emissivity $\varepsilon$ with the Stefan-Boltzmann constant $\sigma_{\mathrm{SB}}$  and the surface temperature $T$ at a specific location on the drum.\cite{Bergman2017} Taking into account thermal radiation, whereby part of the thermal power is emitted, leads to a reduced effective temperature of the drum. It is obvious from equation~(\ref{eq:df}) that a lower average temperature results in a smaller frequency detuning $\Delta f_1$, compared to the case of negligible thermal radiation $\Delta f_2$, as schematically depicted in Fig.~\ref{fig:fig1}(c).

\begin{figure}
\includegraphics{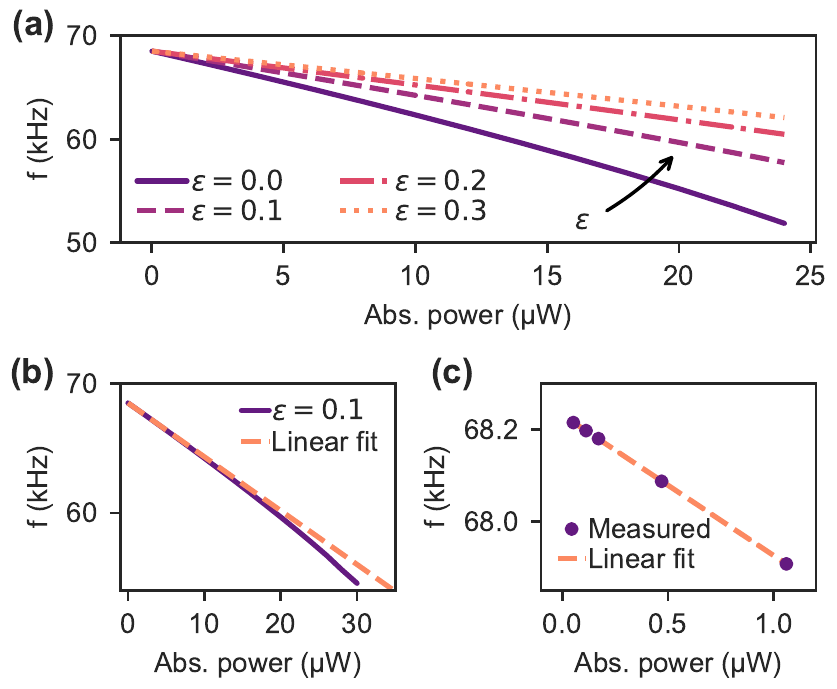}
\caption{\label{fig:fig2} Frequency response of the fundamental mode for an absorbed laser power modeled as a Gaussian beam in the center of a \SI{1}{\mm} SiN drum resonator with a tensile stress of ~\SI{28}{\MPa}. (a) Finite-element method simulation of the fundamental resonator frequency considering different drum surface emissivities, starting from $\varepsilon = 0$ which is equivalent to pure conductive heat transfer inside the drum. (b) Frequency response for an emissivity of $\varepsilon = 0.1$ compared to a linear fit. (c) Measured frequencies for different laser powers of an $L=$\SI{1}{mm} drum with $\sigma=$\SI{28}{\MPa} compared to a linear fit.}
\end{figure}

Fig.~\ref{fig:fig2}(a) shows simulated frequency responses of a \SI{1}{\mm} square drum with different emissivities for an increasing absorbed power $P_{\mathrm{abs}}$. As predicted by the simplified model (\ref{eq:df}), the resonator frequency decreases with increasing absorbed laser power $P_{\mathrm{abs}}$, which corresponds to a rise of the drum's effective temperature. It also shows that the slope of the frequency detuning, and hence the responsivity, becomes smaller when more heat is radiated due to a higher emissivity of the drum.

\begin{figure} 
\includegraphics{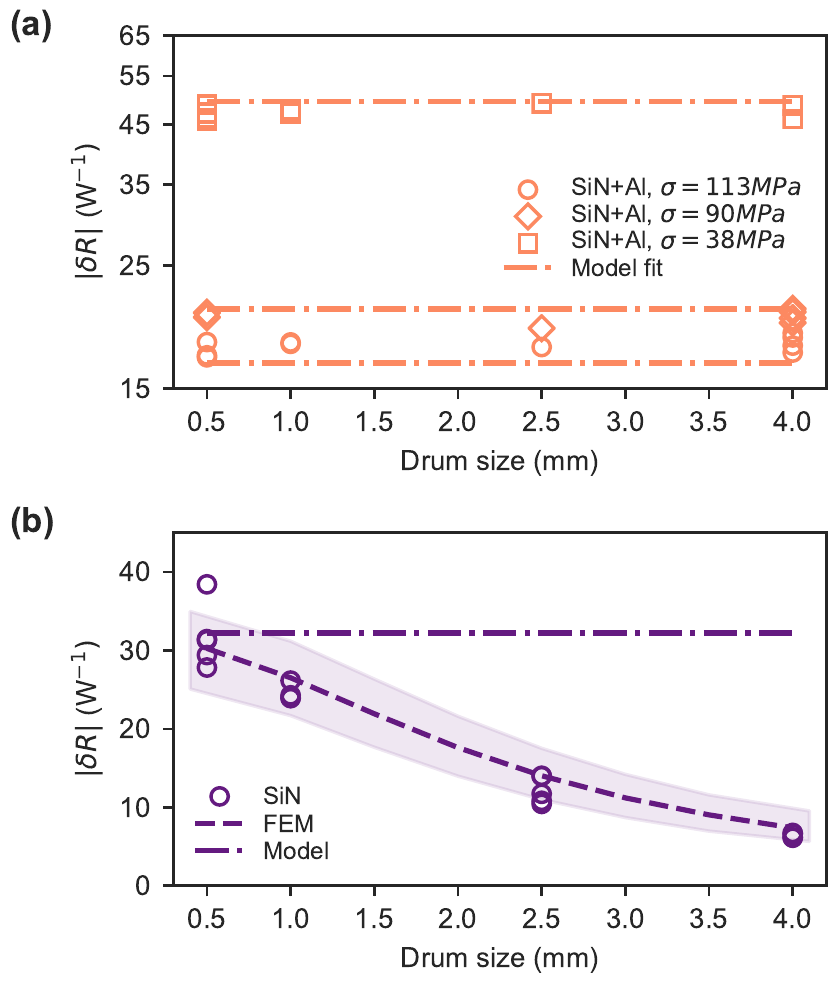}
\caption{\label{fig:fig4} Relative responsivity analysis. Each data point represents the averaged responsivity corresponding to one individual sample, with a total of 27 Al coated and 18 bare SiN samples. (a) Relative responsivity of Al coated SiN drums compared to a fit from the analytical responsivity model for a thermal conduction-dominated heat transfer given in equation~(\ref{eq:responsivity_mbrn}). The fit parameter was found to be $\delta R = \xi/\sigma$ with $\xi \approx 1.88\cdot10^3\,$\si{\MPa\per\W} according to the drum stress $\sigma$. (b) Relative responsivity of bare SiN drums compared to the conduction-dominated model (\ref{eq:responsivity_mbrn}), with $\sigma = 30\,$\si{\MPa}, $\alpha = $\SI{2.2d-6}{\per\K}, and $\kappa = 3\,$\si{\W\per\m\per\K}, and FEM simulations assuming a surface emissivity $\varepsilon = 0.05$, $\alpha=$\SI{2.2d-6}{\per\K},\cite{Toivola2003}. The error band represents the uncertainties in $\kappa$ and $\sigma$ originated for a range from  \numrange{2.7}{3.5}\,\si{\W\per\m\per\K} and \numrange{28}{30}\,\si{\MPa}, respectively.}
\end{figure}

For small changes of temperature $\Delta T$, that is for small $P_{\mathrm{abs}}$, the frequency response (\ref{eq:df}) can be linearized to a good approximation. From Fig.~\ref{fig:fig2}(b) it can be seen that this linear approximation is valid for $P_{\mathrm{abs}}<10\,$\si{\micro\watt}. Considering the applied laser powers and the assumed $A_{\lambda}\approx 0.4\%$ for SiN, which is close to the reported absorbance value of 0.5$\%$\cite{Chien2018} our maximal absorbed power is $P_{\mathrm{abs}}\approx 1\,$\si{\micro\watt}.
Fig.~\ref{fig:fig2}(c) shows that the measured frequency detuning is indeed linear with the absorbed power. Based on this presented method, we measured the relative power responsivity $\delta R = \delta f/P_{\mathrm{abs}}$ for Al coated and bare SiN drum resonators, shown in Fig.~\ref{fig:fig4}(a) and Fig.~\ref{fig:fig4}(b), respectively.

For the case of negligible thermal radiation, an analytical power responsivity model for the fundamental mode has been derived for the case of local heating in the center of a circular drum\cite{Kurek2017} 
\begin{equation}\label{eq:responsivity_mbrn}
\delta R =\frac{\delta f}{P_{\mathrm{abs}}} \approx-\frac{\alpha E}{8 \pi \kappa h \sigma}\left(\frac{2-\nu}{1-\nu}-0.642\right),
\end{equation}
with the Poisson's ratio $\nu$. A comparison to finite element method (FEM) simulations show that the analytical model is a good approximation for square drums.\cite{Kurek2017}
According to (\ref{eq:responsivity_mbrn}), the power responsivity is independent of the lateral drum size. This is the case for the Al coated SiN drums, as can be seen in Fig.~\ref{fig:fig4}(a). Compared to the thermal conductivity of SiN of $\kappa$ = \SI{3}{\W\per\m\per\K},\cite{Ftouni2015} the effective conductivity of the Al coated drums is $\sim$32 times higher; thus, the heat transfer in the Al coated drums seems to be dominated by conduction. The fit of (\ref{eq:responsivity_mbrn}) for the Al coated drums with different effective tensile stress is of good quality.  In contrast, as seen in Fig.~\ref{fig:fig4}(b), the measured responsivities for bare SiN drum resonators decreases with increasing drum size. Even the smallest drums show a deviation from the pure conductive model (\ref{eq:responsivity_mbrn}), suggesting that all measured bare SiN drums are dominated by radiative heat transfer. In order to take radiative heat transfer into account, we used finite element method simulations to model the responsivity of the bare SiN drum resonators. Prior to this step, we measured the absorption spectrum of our SiN drums and calculated their emissivity to be $\varepsilon = 0.05$ (for details see Supplementary Information), which agrees well with values predicted by Zhang et al.\cite{Zhang2020} The simulated responsivities plotted in Fig.~\ref{fig:fig4}(b) follow the measured values with good agreement.

\begin{figure} 
\includegraphics{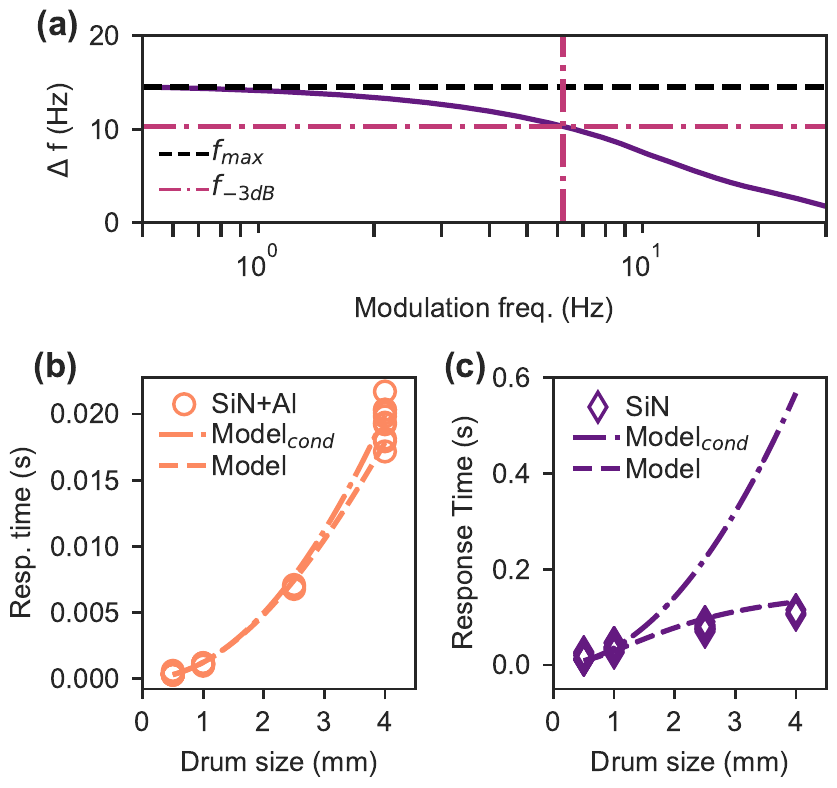}
\caption{\label{fig:fig3} Thermal response time analysis. Each data point corresponds to a measured $\tau$ value of an individual sample with a total of 31 bare SiN and 34 Al coated samples. (a) Measured frequency response of a \SI{1}{\mm} bare SiN drum for an increasing modulation frequency of the heating laser. (b) Response times of Al coated SiN drums of various sizes compared with the analytic model (\ref{eq:tau}) for $\kappa$ = \SI{190}{\W\per\m\per\K},\cite{Feng2009} $\rho$ = \SI{2700}{\kg\per\cubic\metre}, and $C_p$ = \SI{1100}{\J\per\kg\per\K}.\cite{Lugo2016} (c) Response times of bare SiN drum resonators of various sizes compared with the response time model, for $\kappa$ = \SI{3}{\W\per\m\per\K},\cite{Ftouni2015} $\rho$ = \SI{3000}{\kg\per\cubic\metre}, and $C_p$ = \SI{700}{\J\per\kg\per\K}.\cite{Larsen2013ACSNano}}
\end{figure}

Finally, we studied the response time using frequency response measurements for an increasing laser modulation frequency as shown in Fig.~\ref{fig:fig3}(a). By increasing the modulation frequency of the heating laser, the recorded resonance frequency detuning from the phase-locked loop will start to decay. The response time $\tau$ is evaluated for the frequency amplitude at $f_{\mathrm{-3dB}}=f_{\mathrm{max}}/\sqrt{2}$. The extracted response times of Al covered SiN drums are plotted in Fig.~\ref{fig:fig3}(b). Compared to the measurements of bare SiN drums, shown in Fig.~\ref{fig:fig3}(c), the Al coated drums show a significantly faster response time, due to their higher thermal conductivity and hence faster thermalization. Besides the magnitude, the scaling of $\tau$ with lateral size is notably different for the two drum types.

Our theoretical model (see Supplementary Information for the derivation) for the response time of square drums made of multiple layers, considering both thermal conduction and thermal radiation, yields
\begin{equation}\label{eq:tau}
\tau=\left(\underbrace{\frac{2\pi^{2}}{L^{2}}\frac{\sum_{i}h_{i}\kappa_{i}}%
{\sum_{i}h_{i} \rho_{i} C_{p_i}}}_{\text{Conduction}}+
\underbrace{\frac{8\epsilon\sigma_{\mathrm{SB}}T_{0}^{3}}{\sum_{i}h_{i} \rho_{i} C_{p_i}}}_{\text{Radiation}}\right)^{-1},
\end{equation} 
with the mass density $\rho$, the specific heat capacity $C_p$ for each layer $i$.  The model (\ref{eq:tau}) is in excellent agreement with the measured response times for both drum types, as seen from Fig.~\ref{fig:fig3}(b)\&(c). In order to dissect the dominating heat transfer mechanism at play, we additionally plotted the model taking into account only heat conduction. In that case, (\ref{eq:tau}) predicts a quadratic scaling with drum size $L$, and is a good approximation in the case for the conduction-dominated Al coated drums, as seen in Fig.~\ref{fig:fig3}(b). In comparison, only the full model (\ref{eq:tau}),  taking into account both conduction and radiation, predicts the measured response times accurately, as seen in Fig.~\ref{fig:fig3}(b).    
This is another clear sign that heat transfer in Al coated SiN drums is dominated by conduction in contrast to the bare SiN drums that are radiation-limited. The higher radiative heat loss for bigger drums leads to a reduced response time that levels off, compared to what would be expected from heat conduction alone.

In conclusion, it has been demonstrated that the thermal frequency response of nanomechanical SiN drum resonators is significantly affected by radiative heat transfer. This results in a size-dependent responsivity of bare SiN drums in contrast to Al coated resonators where thermal conduction dominates and no size-dependency on responsivity was observed.
The dominating heat transfer mechanism of the resonator is also reflected in response time measurements. The Al-covered drums are dominated by thermal conduction and show a significantly faster response time due to the high thermal conductivity of the metal layer.  
An analytical model for this case corresponds well with the measurements and show that the response time scales quadratically with the drum size. For bare SiN drums, where thermal radiation plays a non-negligible role, measurements also showed good agreement with the response time model, predicting faster response times due to the additional radiative heat loss. Larger drums are more affected by radiation, exhibiting an even higher deviation for response times compared to the case of a thermal conduction-dominated heat transfer. The obtained results show the significance of thermal radiation in finding the optimal performance of SiN drum resonators for specific sensor applications and fundamental research.

\begin{acknowledgments}
The authors wish to thank Johannes Hiesberger, Sophia Ewert, Patrick Meyer, and Michael Buchholz for their support with the sample fabrication as well as Hendrik Kähler and Miao-Hsuan Chien for many fruitful discussions. We would also like to thank Dr. Pavel Grinchuk of HMTI, Belarus for his support. This work is supported by the European Research Council under the European Unions Horizon 2020 research and innovation program (Grant Agreement-716087-PLASMECS) and (Grant Agreement-875518-NIRD). We further acknowledge funding from Invisible-Light Labs GmbH.
\end{acknowledgments}

\bibliography{literature, silvanbib,literature_support}

\begin{thebibliography}{36}%
\makeatletter
\providecommand \@ifxundefined [1]{%
 \@ifx{#1\undefined}
}%
\providecommand \@ifnum [1]{%
 \ifnum #1\expandafter \@firstoftwo
 \else \expandafter \@secondoftwo
 \fi
}%
\providecommand \@ifx [1]{%
 \ifx #1\expandafter \@firstoftwo
 \else \expandafter \@secondoftwo
 \fi
}%
\providecommand \natexlab [1]{#1}%
\providecommand \enquote  [1]{``#1''}%
\providecommand \bibnamefont  [1]{#1}%
\providecommand \bibfnamefont [1]{#1}%
\providecommand \citenamefont [1]{#1}%
\providecommand \href@noop [0]{\@secondoftwo}%
\providecommand \href [0]{\begingroup \@sanitize@url \@href}%
\providecommand \@href[1]{\@@startlink{#1}\@@href}%
\providecommand \@@href[1]{\endgroup#1\@@endlink}%
\providecommand \@sanitize@url [0]{\catcode `\\12\catcode `\$12\catcode
  `\&12\catcode `\#12\catcode `\^12\catcode `\_12\catcode `\%12\relax}%
\providecommand \@@startlink[1]{}%
\providecommand \@@endlink[0]{}%
\providecommand \url  [0]{\begingroup\@sanitize@url \@url }%
\providecommand \@url [1]{\endgroup\@href {#1}{\urlprefix }}%
\providecommand \urlprefix  [0]{URL }%
\providecommand \Eprint [0]{\href }%
\providecommand \doibase [0]{http://dx.doi.org/}%
\providecommand \selectlanguage [0]{\@gobble}%
\providecommand \bibinfo  [0]{\@secondoftwo}%
\providecommand \bibfield  [0]{\@secondoftwo}%
\providecommand \translation [1]{[#1]}%
\providecommand \BibitemOpen [0]{}%
\providecommand \bibitemStop [0]{}%
\providecommand \bibitemNoStop [0]{.\EOS\space}%
\providecommand \EOS [0]{\spacefactor3000\relax}%
\providecommand \BibitemShut  [1]{\csname bibitem#1\endcsname}%
\let\auto@bib@innerbib\@empty
\bibitem [{\citenamefont {Schmid}, \citenamefont {Villanueva},\ and\
  \citenamefont {Roukes}(2016)}]{Schmid2016}%
  \BibitemOpen
  \bibfield  {author} {\bibinfo {author} {\bibfnamefont {S.}~\bibnamefont
  {Schmid}}, \bibinfo {author} {\bibfnamefont {L.~G.}\ \bibnamefont
  {Villanueva}}, \ and\ \bibinfo {author} {\bibfnamefont {M.~L.}\ \bibnamefont
  {Roukes}},\ }\href {\doibase 10.1007/978-3-319-28691-4} {\emph {\bibinfo
  {title} {Fundamentals of Nanomechanical Resonators}}}\ (\bibinfo  {publisher}
  {Springer International Publishing},\ \bibinfo {address} {Cham},\ \bibinfo
  {year} {2016})\ pp.\ \bibinfo {pages} {1--175}\BibitemShut {NoStop}%
\bibitem [{\citenamefont {Verbridge}\ \emph {et~al.}(2006)\citenamefont
  {Verbridge}, \citenamefont {Parpia}, \citenamefont {Reichenbach},
  \citenamefont {Bellan},\ and\ \citenamefont {Craighead}}]{Verbridge2006}%
  \BibitemOpen
  \bibfield  {author} {\bibinfo {author} {\bibfnamefont {S.~S.}\ \bibnamefont
  {Verbridge}}, \bibinfo {author} {\bibfnamefont {J.~M.}\ \bibnamefont
  {Parpia}}, \bibinfo {author} {\bibfnamefont {R.~B.}\ \bibnamefont
  {Reichenbach}}, \bibinfo {author} {\bibfnamefont {L.~M.}\ \bibnamefont
  {Bellan}}, \ and\ \bibinfo {author} {\bibfnamefont {H.~G.}\ \bibnamefont
  {Craighead}},\ }\bibfield  {title} {\enquote {\bibinfo {title} {High quality
  factor resonance at room temperature with nanostrings under high tensile
  stress},}\ }\href {\doibase 10.1063/1.2204829} {\bibfield  {journal}
  {\bibinfo  {journal} {Journal of Applied Physics}\ }\textbf {\bibinfo
  {volume} {99}},\ \bibinfo {pages} {124304} (\bibinfo {year}
  {2006})}\BibitemShut {NoStop}%
\bibitem [{\citenamefont {Schmid}\ \emph {et~al.}(2011)\citenamefont {Schmid},
  \citenamefont {Jensen}, \citenamefont {Nielsen},\ and\ \citenamefont
  {Boisen}}]{Schmid2011}%
  \BibitemOpen
  \bibfield  {author} {\bibinfo {author} {\bibfnamefont {S.}~\bibnamefont
  {Schmid}}, \bibinfo {author} {\bibfnamefont {K.~D.}\ \bibnamefont {Jensen}},
  \bibinfo {author} {\bibfnamefont {K.~H.}\ \bibnamefont {Nielsen}}, \ and\
  \bibinfo {author} {\bibfnamefont {A.}~\bibnamefont {Boisen}},\ }\bibfield
  {title} {\enquote {\bibinfo {title} {Damping mechanisms in high-$q$ micro and
  nanomechanical string resonators},}\ }\href {\doibase
  10.1103/PhysRevB.84.165307} {\bibfield  {journal} {\bibinfo  {journal} {Phys.
  Rev. B}\ }\textbf {\bibinfo {volume} {84}},\ \bibinfo {pages} {165307}
  (\bibinfo {year} {2011})}\BibitemShut {NoStop}%
\bibitem [{\citenamefont {Zwickl}\ \emph {et~al.}(2008)\citenamefont {Zwickl},
  \citenamefont {Shanks}, \citenamefont {Jayich}, \citenamefont {Yang},
  \citenamefont {Jayich}, \citenamefont {Thompson},\ and\ \citenamefont
  {Harris}}]{Zwickl2008}%
  \BibitemOpen
  \bibfield  {author} {\bibinfo {author} {\bibfnamefont {B.~M.}\ \bibnamefont
  {Zwickl}}, \bibinfo {author} {\bibfnamefont {W.~E.}\ \bibnamefont {Shanks}},
  \bibinfo {author} {\bibfnamefont {A.~M.}\ \bibnamefont {Jayich}}, \bibinfo
  {author} {\bibfnamefont {C.}~\bibnamefont {Yang}}, \bibinfo {author}
  {\bibfnamefont {A.~C.}\ \bibnamefont {Jayich}}, \bibinfo {author}
  {\bibfnamefont {J.~D.}\ \bibnamefont {Thompson}}, \ and\ \bibinfo {author}
  {\bibfnamefont {J.~G.}\ \bibnamefont {Harris}},\ }\bibfield  {title}
  {\enquote {\bibinfo {title} {{High quality mechanical and optical properties
  of commercial silicon nitride membranes}},}\ }\href {\doibase
  10.1063/1.2884191} {\bibfield  {journal} {\bibinfo  {journal} {Applied
  Physics Letters}\ }\textbf {\bibinfo {volume} {92}},\ \bibinfo {pages}
  {2006--2009} (\bibinfo {year} {2008})}\BibitemShut {NoStop}%
\bibitem [{\citenamefont {Wilson}\ \emph {et~al.}(2009)\citenamefont {Wilson},
  \citenamefont {Regal}, \citenamefont {Papp},\ and\ \citenamefont
  {Kimble}}]{Wilson2009}%
  \BibitemOpen
  \bibfield  {author} {\bibinfo {author} {\bibfnamefont {D.~J.}\ \bibnamefont
  {Wilson}}, \bibinfo {author} {\bibfnamefont {C.~A.}\ \bibnamefont {Regal}},
  \bibinfo {author} {\bibfnamefont {S.~B.}\ \bibnamefont {Papp}}, \ and\
  \bibinfo {author} {\bibfnamefont {H.~J.}\ \bibnamefont {Kimble}},\ }\bibfield
   {title} {\enquote {\bibinfo {title} {Cavity optomechanics with
  stoichiometric sin films},}\ }\href {\doibase 10.1103/PhysRevLett.103.207204}
  {\bibfield  {journal} {\bibinfo  {journal} {Phys. Rev. Lett.}\ }\textbf
  {\bibinfo {volume} {103}},\ \bibinfo {pages} {207204} (\bibinfo {year}
  {2009})}\BibitemShut {NoStop}%
\bibitem [{\citenamefont {Yu}, \citenamefont {Purdy},\ and\ \citenamefont
  {Regal}(2012)}]{Yu2012}%
  \BibitemOpen
  \bibfield  {author} {\bibinfo {author} {\bibfnamefont {P.-L.}\ \bibnamefont
  {Yu}}, \bibinfo {author} {\bibfnamefont {T.~P.}\ \bibnamefont {Purdy}}, \
  and\ \bibinfo {author} {\bibfnamefont {C.~A.}\ \bibnamefont {Regal}},\
  }\bibfield  {title} {\enquote {\bibinfo {title} {{Control of Material Damping
  in High-Q Membrane Microresonators}},}\ }\href {\doibase
  10.1103/PhysRevLett.108.083603} {\bibfield  {journal} {\bibinfo  {journal}
  {Physical Review Letters}\ }\textbf {\bibinfo {volume} {108}},\ \bibinfo
  {pages} {083603} (\bibinfo {year} {2012})}\BibitemShut {NoStop}%
\bibitem [{\citenamefont {Thompson}\ \emph {et~al.}(2008)\citenamefont
  {Thompson}, \citenamefont {Zwickl}, \citenamefont {Jayich}, \citenamefont
  {Marquardt}, \citenamefont {Girvin},\ and\ \citenamefont
  {Harris}}]{thompson2008strong}%
  \BibitemOpen
  \bibfield  {author} {\bibinfo {author} {\bibfnamefont {J.~D.}\ \bibnamefont
  {Thompson}}, \bibinfo {author} {\bibfnamefont {B.~M.}\ \bibnamefont
  {Zwickl}}, \bibinfo {author} {\bibfnamefont {A.~M.}\ \bibnamefont {Jayich}},
  \bibinfo {author} {\bibfnamefont {F.}~\bibnamefont {Marquardt}}, \bibinfo
  {author} {\bibfnamefont {S.~M.}\ \bibnamefont {Girvin}}, \ and\ \bibinfo
  {author} {\bibfnamefont {J.~G.~E.}\ \bibnamefont {Harris}},\ }\bibfield
  {title} {\enquote {\bibinfo {title} {{Strong dispersive coupling of a
  high-finesse cavity to a micromechanical membrane}},}\ }\href@noop {}
  {\bibfield  {journal} {\bibinfo  {journal} {Nature}\ }\textbf {\bibinfo
  {volume} {452}},\ \bibinfo {pages} {72--75} (\bibinfo {year}
  {2008})}\BibitemShut {NoStop}%
\bibitem [{\citenamefont {Purdy}, \citenamefont {Peterson},\ and\ \citenamefont
  {Regal}(2013)}]{Purdy2013}%
  \BibitemOpen
  \bibfield  {author} {\bibinfo {author} {\bibfnamefont {T.~P.}\ \bibnamefont
  {Purdy}}, \bibinfo {author} {\bibfnamefont {R.~W.}\ \bibnamefont {Peterson}},
  \ and\ \bibinfo {author} {\bibfnamefont {C.~A.}\ \bibnamefont {Regal}},\
  }\bibfield  {title} {\enquote {\bibinfo {title} {Observation of radiation
  pressure shot noise on a macroscopic object},}\ }\href {\doibase
  10.1126/science.1231282} {\bibfield  {journal} {\bibinfo  {journal}
  {Science}\ }\textbf {\bibinfo {volume} {339}},\ \bibinfo {pages} {801--804}
  (\bibinfo {year} {2013})},\ \Eprint
  {http://arxiv.org/abs/https://science.sciencemag.org/content/339/6121/801.full.pdf}
  {https://science.sciencemag.org/content/339/6121/801.full.pdf} \BibitemShut
  {NoStop}%
\bibitem [{\citenamefont {Bagci}\ \emph {et~al.}(2014)\citenamefont {Bagci},
  \citenamefont {Simonsen}, \citenamefont {Schmid}, \citenamefont {Villanueva},
  \citenamefont {Zeuthen}, \citenamefont {Appel}, \citenamefont {Taylor},
  \citenamefont {S{\o}rensen}, \citenamefont {Usami}, \citenamefont
  {Schliesser},\ and\ \citenamefont {Polzik}}]{Bagci2014}%
  \BibitemOpen
  \bibfield  {author} {\bibinfo {author} {\bibfnamefont {T.}~\bibnamefont
  {Bagci}}, \bibinfo {author} {\bibfnamefont {A.}~\bibnamefont {Simonsen}},
  \bibinfo {author} {\bibfnamefont {S.}~\bibnamefont {Schmid}}, \bibinfo
  {author} {\bibfnamefont {L.~G.}\ \bibnamefont {Villanueva}}, \bibinfo
  {author} {\bibfnamefont {E.}~\bibnamefont {Zeuthen}}, \bibinfo {author}
  {\bibfnamefont {J.}~\bibnamefont {Appel}}, \bibinfo {author} {\bibfnamefont
  {J.~M.}\ \bibnamefont {Taylor}}, \bibinfo {author} {\bibfnamefont
  {A.}~\bibnamefont {S{\o}rensen}}, \bibinfo {author} {\bibfnamefont
  {K.}~\bibnamefont {Usami}}, \bibinfo {author} {\bibfnamefont
  {A.}~\bibnamefont {Schliesser}}, \ and\ \bibinfo {author} {\bibfnamefont
  {E.~S.}\ \bibnamefont {Polzik}},\ }\bibfield  {title} {\enquote {\bibinfo
  {title} {{Optical detection of radio waves through a nanomechanical
  transducer}},}\ }\href {\doibase 10.1038/nature13029} {\bibfield  {journal}
  {\bibinfo  {journal} {Nature}\ }\textbf {\bibinfo {volume} {507}},\ \bibinfo
  {pages} {81--85} (\bibinfo {year} {2014})}\BibitemShut {NoStop}%
\bibitem [{\citenamefont {Andrews}\ \emph {et~al.}(2014)\citenamefont
  {Andrews}, \citenamefont {Peterson}, \citenamefont {Purdy}, \citenamefont
  {Cicak}, \citenamefont {Simmonds}, \citenamefont {Regal},\ and\ \citenamefont
  {Lehnert}}]{Andrews2014}%
  \BibitemOpen
  \bibfield  {author} {\bibinfo {author} {\bibfnamefont {R.~W.}\ \bibnamefont
  {Andrews}}, \bibinfo {author} {\bibfnamefont {R.~W.}\ \bibnamefont
  {Peterson}}, \bibinfo {author} {\bibfnamefont {T.~P.}\ \bibnamefont {Purdy}},
  \bibinfo {author} {\bibfnamefont {K.}~\bibnamefont {Cicak}}, \bibinfo
  {author} {\bibfnamefont {R.~W.}\ \bibnamefont {Simmonds}}, \bibinfo {author}
  {\bibfnamefont {C.~A.}\ \bibnamefont {Regal}}, \ and\ \bibinfo {author}
  {\bibfnamefont {K.~W.}\ \bibnamefont {Lehnert}},\ }\bibfield  {title}
  {\enquote {\bibinfo {title} {{Bidirectional and efficient conversion between
  microwave and optical light}},}\ }\href {\doibase 10.1038/NPHYS2911}
  {\bibfield  {journal} {\bibinfo  {journal} {Nature Physics}\ }\textbf
  {\bibinfo {volume} {10}},\ \bibinfo {pages} {321--326} (\bibinfo {year}
  {2014})}\BibitemShut {NoStop}%
\bibitem [{\citenamefont {Pluchar}\ \emph {et~al.}(2020)\citenamefont
  {Pluchar}, \citenamefont {Agrawal}, \citenamefont {Schenk},\ and\
  \citenamefont {Wilson}}]{pluchar2020cavityfree}%
  \BibitemOpen
  \bibfield  {author} {\bibinfo {author} {\bibfnamefont {C.~M.}\ \bibnamefont
  {Pluchar}}, \bibinfo {author} {\bibfnamefont {A.}~\bibnamefont {Agrawal}},
  \bibinfo {author} {\bibfnamefont {E.}~\bibnamefont {Schenk}}, \ and\ \bibinfo
  {author} {\bibfnamefont {D.~J.}\ \bibnamefont {Wilson}},\ }\href@noop {}
  {\enquote {\bibinfo {title} {Towards cavity-free ground state cooling of an
  acoustic-frequency silicon nitride membrane},}\ } (\bibinfo {year} {2020}),\
  \Eprint {http://arxiv.org/abs/2004.13187} {arXiv:2004.13187 [quant-ph]}
  \BibitemShut {NoStop}%
\bibitem [{\citenamefont {Norte}, \citenamefont {Moura},\ and\ \citenamefont
  {Gr\"oblacher}(2016)}]{PhysRevLett.116.147202}%
  \BibitemOpen
  \bibfield  {author} {\bibinfo {author} {\bibfnamefont {R.~A.}\ \bibnamefont
  {Norte}}, \bibinfo {author} {\bibfnamefont {J.~P.}\ \bibnamefont {Moura}}, \
  and\ \bibinfo {author} {\bibfnamefont {S.}~\bibnamefont {Gr\"oblacher}},\
  }\bibfield  {title} {\enquote {\bibinfo {title} {Mechanical resonators for
  quantum optomechanics experiments at room temperature},}\ }\href {\doibase
  10.1103/PhysRevLett.116.147202} {\bibfield  {journal} {\bibinfo  {journal}
  {Phys. Rev. Lett.}\ }\textbf {\bibinfo {volume} {116}},\ \bibinfo {pages}
  {147202} (\bibinfo {year} {2016})}\BibitemShut {NoStop}%
\bibitem [{\citenamefont {Reinhardt}\ \emph {et~al.}(2016)\citenamefont
  {Reinhardt}, \citenamefont {M\"uller}, \citenamefont {Bourassa},\ and\
  \citenamefont {Sankey}}]{PhysRevX.6.021001}%
  \BibitemOpen
  \bibfield  {author} {\bibinfo {author} {\bibfnamefont {C.}~\bibnamefont
  {Reinhardt}}, \bibinfo {author} {\bibfnamefont {T.}~\bibnamefont {M\"uller}},
  \bibinfo {author} {\bibfnamefont {A.}~\bibnamefont {Bourassa}}, \ and\
  \bibinfo {author} {\bibfnamefont {J.~C.}\ \bibnamefont {Sankey}},\ }\bibfield
   {title} {\enquote {\bibinfo {title} {Ultralow-noise sin trampoline
  resonators for sensing and optomechanics},}\ }\href {\doibase
  10.1103/PhysRevX.6.021001} {\bibfield  {journal} {\bibinfo  {journal} {Phys.
  Rev. X}\ }\textbf {\bibinfo {volume} {6}},\ \bibinfo {pages} {021001}
  (\bibinfo {year} {2016})}\BibitemShut {NoStop}%
\bibitem [{\citenamefont {Tsaturyan}\ \emph {et~al.}(2017)\citenamefont
  {Tsaturyan}, \citenamefont {Barg}, \citenamefont {Polzik},\ and\
  \citenamefont {Schliesser}}]{Tsaturyan2016a}%
  \BibitemOpen
  \bibfield  {author} {\bibinfo {author} {\bibfnamefont {Y.}~\bibnamefont
  {Tsaturyan}}, \bibinfo {author} {\bibfnamefont {A.}~\bibnamefont {Barg}},
  \bibinfo {author} {\bibfnamefont {E.~S.}\ \bibnamefont {Polzik}}, \ and\
  \bibinfo {author} {\bibfnamefont {A.}~\bibnamefont {Schliesser}},\ }\bibfield
   {title} {\enquote {\bibinfo {title} {{Ultracoherent nanomechanical
  resonators via soft clamping and dissipation dilution}},}\ }\href {\doibase
  10.1038/nnano.2017.101} {\bibfield  {journal} {\bibinfo  {journal} {Nature
  Nanotechnology}\ }\textbf {\bibinfo {volume} {12}},\ \bibinfo {pages}
  {776--783} (\bibinfo {year} {2017})},\ \Eprint
  {http://arxiv.org/abs/1608.00937} {arXiv:1608.00937} \BibitemShut {NoStop}%
\bibitem [{\citenamefont {Ghadimi}\ \emph {et~al.}(2018)\citenamefont
  {Ghadimi}, \citenamefont {Fedorov}, \citenamefont {Engelsen}, \citenamefont
  {Bereyhi}, \citenamefont {Schilling}, \citenamefont {Wilson},\ and\
  \citenamefont {Kippenberg}}]{Ghadimi764}%
  \BibitemOpen
  \bibfield  {author} {\bibinfo {author} {\bibfnamefont {A.~H.}\ \bibnamefont
  {Ghadimi}}, \bibinfo {author} {\bibfnamefont {S.~A.}\ \bibnamefont
  {Fedorov}}, \bibinfo {author} {\bibfnamefont {N.~J.}\ \bibnamefont
  {Engelsen}}, \bibinfo {author} {\bibfnamefont {M.~J.}\ \bibnamefont
  {Bereyhi}}, \bibinfo {author} {\bibfnamefont {R.}~\bibnamefont {Schilling}},
  \bibinfo {author} {\bibfnamefont {D.~J.}\ \bibnamefont {Wilson}}, \ and\
  \bibinfo {author} {\bibfnamefont {T.~J.}\ \bibnamefont {Kippenberg}},\
  }\bibfield  {title} {\enquote {\bibinfo {title} {Elastic strain engineering
  for ultralow mechanical dissipation},}\ }\href {\doibase
  10.1126/science.aar6939} {\bibfield  {journal} {\bibinfo  {journal}
  {Science}\ }\textbf {\bibinfo {volume} {360}},\ \bibinfo {pages} {764--768}
  (\bibinfo {year} {2018})},\ \Eprint
  {http://arxiv.org/abs/https://science.sciencemag.org/content/360/6390/764.full.pdf}
  {https://science.sciencemag.org/content/360/6390/764.full.pdf} \BibitemShut
  {NoStop}%
\bibitem [{\citenamefont {Reetz}\ \emph {et~al.}(2019)\citenamefont {Reetz},
  \citenamefont {Fischer}, \citenamefont {Assump\ifmmode
  \mbox{\c{c}}\else~\c{c}\fi{}\ ao}, \citenamefont {McNally}, \citenamefont
  {Burns}, \citenamefont {Sankey},\ and\ \citenamefont {Regal}}]{Reetz2019}%
  \BibitemOpen
  \bibfield  {author} {\bibinfo {author} {\bibfnamefont {C.}~\bibnamefont
  {Reetz}}, \bibinfo {author} {\bibfnamefont {R.}~\bibnamefont {Fischer}},
  \bibinfo {author} {\bibfnamefont {G.}~\bibnamefont {Assump\ifmmode
  \mbox{\c{c}}\else~\c{c}\fi{}\ ao}}, \bibinfo {author} {\bibfnamefont
  {D.}~\bibnamefont {McNally}}, \bibinfo {author} {\bibfnamefont
  {P.}~\bibnamefont {Burns}}, \bibinfo {author} {\bibfnamefont
  {J.}~\bibnamefont {Sankey}}, \ and\ \bibinfo {author} {\bibfnamefont
  {C.}~\bibnamefont {Regal}},\ }\bibfield  {title} {\enquote {\bibinfo {title}
  {Analysis of membrane phononic crystals with wide band gaps and low-mass
  defects},}\ }\href {\doibase 10.1103/PhysRevApplied.12.044027} {\bibfield
  {journal} {\bibinfo  {journal} {Phys. Rev. Applied}\ }\textbf {\bibinfo
  {volume} {12}},\ \bibinfo {pages} {044027} (\bibinfo {year}
  {2019})}\BibitemShut {NoStop}%
\bibitem [{\citenamefont {Friedrich}\ \emph {et~al.}(2011)\citenamefont
  {Friedrich}, \citenamefont {Kaufer}, \citenamefont {Westphal}, \citenamefont
  {Yamamoto}, \citenamefont {Sawadsky}, \citenamefont {{Ya Khalili}},
  \citenamefont {Danilishin}, \citenamefont {Go{\ss}ler}, \citenamefont
  {Danzmann},\ and\ \citenamefont {Schnabel}}]{Friedrich2011}%
  \BibitemOpen
  \bibfield  {author} {\bibinfo {author} {\bibfnamefont {D.}~\bibnamefont
  {Friedrich}}, \bibinfo {author} {\bibfnamefont {H.}~\bibnamefont {Kaufer}},
  \bibinfo {author} {\bibfnamefont {T.}~\bibnamefont {Westphal}}, \bibinfo
  {author} {\bibfnamefont {K.}~\bibnamefont {Yamamoto}}, \bibinfo {author}
  {\bibfnamefont {A.}~\bibnamefont {Sawadsky}}, \bibinfo {author}
  {\bibfnamefont {F.}~\bibnamefont {{Ya Khalili}}}, \bibinfo {author}
  {\bibfnamefont {S.~L.}\ \bibnamefont {Danilishin}}, \bibinfo {author}
  {\bibfnamefont {S.}~\bibnamefont {Go{\ss}ler}}, \bibinfo {author}
  {\bibfnamefont {K.}~\bibnamefont {Danzmann}}, \ and\ \bibinfo {author}
  {\bibfnamefont {R.}~\bibnamefont {Schnabel}},\ }\bibfield  {title} {\enquote
  {\bibinfo {title} {{Laser interferometry with translucent and absorbing
  mechanical oscillators}},}\ }\href {\doibase 10.1088/1367-2630/13/9/093017}
  {\bibfield  {journal} {\bibinfo  {journal} {New Journal of Physics}\ }\textbf
  {\bibinfo {volume} {13}} (\bibinfo {year} {2011}),\
  10.1088/1367-2630/13/9/093017}\BibitemShut {NoStop}%
\bibitem [{\citenamefont {J{\"{o}}ckel}\ \emph {et~al.}(2011)\citenamefont
  {J{\"{o}}ckel}, \citenamefont {Rakher}, \citenamefont {Korppi}, \citenamefont
  {Camerer}, \citenamefont {Hunger}, \citenamefont {Mader},\ and\ \citenamefont
  {Treutlein}}]{Jockel2011}%
  \BibitemOpen
  \bibfield  {author} {\bibinfo {author} {\bibfnamefont {A.}~\bibnamefont
  {J{\"{o}}ckel}}, \bibinfo {author} {\bibfnamefont {M.~T.}\ \bibnamefont
  {Rakher}}, \bibinfo {author} {\bibfnamefont {M.}~\bibnamefont {Korppi}},
  \bibinfo {author} {\bibfnamefont {S.}~\bibnamefont {Camerer}}, \bibinfo
  {author} {\bibfnamefont {D.}~\bibnamefont {Hunger}}, \bibinfo {author}
  {\bibfnamefont {M.}~\bibnamefont {Mader}}, \ and\ \bibinfo {author}
  {\bibfnamefont {P.}~\bibnamefont {Treutlein}},\ }\bibfield  {title} {\enquote
  {\bibinfo {title} {{Spectroscopy of mechanical dissipation in
  micro-mechanical membranes}},}\ }\href {\doibase 10.1063/1.3646914}
  {\bibfield  {journal} {\bibinfo  {journal} {Applied Physics Letters}\
  }\textbf {\bibinfo {volume} {99}} (\bibinfo {year} {2011}),\
  10.1063/1.3646914}\BibitemShut {NoStop}%
\bibitem [{\citenamefont {Yamada}\ \emph {et~al.}(2013)\citenamefont {Yamada},
  \citenamefont {Schmid}, \citenamefont {Larsen}, \citenamefont {Hansen},\ and\
  \citenamefont {Boisen}}]{Yamada2013}%
  \BibitemOpen
  \bibfield  {author} {\bibinfo {author} {\bibfnamefont {S.}~\bibnamefont
  {Yamada}}, \bibinfo {author} {\bibfnamefont {S.}~\bibnamefont {Schmid}},
  \bibinfo {author} {\bibfnamefont {T.}~\bibnamefont {Larsen}}, \bibinfo
  {author} {\bibfnamefont {O.}~\bibnamefont {Hansen}}, \ and\ \bibinfo {author}
  {\bibfnamefont {A.}~\bibnamefont {Boisen}},\ }\bibfield  {title} {\enquote
  {\bibinfo {title} {{Photothermal infrared spectroscopy of airborne samples
  with mechanical string resonators}},}\ }\href {\doibase 10.1021/ac402585e}
  {\bibfield  {journal} {\bibinfo  {journal} {Analytical Chemistry}\ }\textbf
  {\bibinfo {volume} {85}},\ \bibinfo {pages} {10531--10535} (\bibinfo {year}
  {2013})}\BibitemShut {NoStop}%
\bibitem [{\citenamefont {Biswas}\ \emph {et~al.}(2014)\citenamefont {Biswas},
  \citenamefont {Miriyala}, \citenamefont {Doolin}, \citenamefont {Liu},
  \citenamefont {Thundat},\ and\ \citenamefont {Davis}}]{Biswas2014}%
  \BibitemOpen
  \bibfield  {author} {\bibinfo {author} {\bibfnamefont {T.~S.}\ \bibnamefont
  {Biswas}}, \bibinfo {author} {\bibfnamefont {N.}~\bibnamefont {Miriyala}},
  \bibinfo {author} {\bibfnamefont {C.}~\bibnamefont {Doolin}}, \bibinfo
  {author} {\bibfnamefont {X.}~\bibnamefont {Liu}}, \bibinfo {author}
  {\bibfnamefont {T.}~\bibnamefont {Thundat}}, \ and\ \bibinfo {author}
  {\bibfnamefont {J.~P.}\ \bibnamefont {Davis}},\ }\bibfield  {title} {\enquote
  {\bibinfo {title} {{Femtogram-Scale Photothermal Spectroscopy of Explosive
  Molecules on Nanostrings}},}\ }\href@noop {} {\bibfield  {journal} {\bibinfo
  {journal} {Analytical chemistry}\ }\textbf {\bibinfo {volume} {86}},\
  \bibinfo {pages} {11368--11372} (\bibinfo {year} {2014})}\BibitemShut
  {NoStop}%
\bibitem [{\citenamefont {Andersen}\ \emph {et~al.}(2016)\citenamefont
  {Andersen}, \citenamefont {Yamada}, \citenamefont {{Kumar E.K.}},
  \citenamefont {Andresen}, \citenamefont {Boisen},\ and\ \citenamefont
  {Schmid}}]{Andersen2016}%
  \BibitemOpen
  \bibfield  {author} {\bibinfo {author} {\bibfnamefont {A.~J.}\ \bibnamefont
  {Andersen}}, \bibinfo {author} {\bibfnamefont {S.}~\bibnamefont {Yamada}},
  \bibinfo {author} {\bibfnamefont {P.}~\bibnamefont {{Kumar E.K.}}}, \bibinfo
  {author} {\bibfnamefont {T.~L.}\ \bibnamefont {Andresen}}, \bibinfo {author}
  {\bibfnamefont {A.}~\bibnamefont {Boisen}}, \ and\ \bibinfo {author}
  {\bibfnamefont {S.}~\bibnamefont {Schmid}},\ }\bibfield  {title} {\enquote
  {\bibinfo {title} {{Nanomechanical IR spectroscopy for fast analysis of
  liquid-dispersed engineered nanomaterials}},}\ }\href {\doibase
  10.1016/j.snb.2016.04.002} {\bibfield  {journal} {\bibinfo  {journal}
  {Sensors and Actuators B: Chemical}\ }\textbf {\bibinfo {volume} {233}},\
  \bibinfo {pages} {667--673} (\bibinfo {year} {2016})}\BibitemShut {NoStop}%
\bibitem [{\citenamefont {Kurek}\ \emph {et~al.}(2017)\citenamefont {Kurek},
  \citenamefont {Carnoy}, \citenamefont {Larsen}, \citenamefont {Nielsen},
  \citenamefont {Hansen}, \citenamefont {Rades}, \citenamefont {Schmid},\ and\
  \citenamefont {Boisen}}]{Kurek2017}%
  \BibitemOpen
  \bibfield  {author} {\bibinfo {author} {\bibfnamefont {M.}~\bibnamefont
  {Kurek}}, \bibinfo {author} {\bibfnamefont {M.}~\bibnamefont {Carnoy}},
  \bibinfo {author} {\bibfnamefont {P.~E.}\ \bibnamefont {Larsen}}, \bibinfo
  {author} {\bibfnamefont {L.~H.}\ \bibnamefont {Nielsen}}, \bibinfo {author}
  {\bibfnamefont {O.}~\bibnamefont {Hansen}}, \bibinfo {author} {\bibfnamefont
  {T.}~\bibnamefont {Rades}}, \bibinfo {author} {\bibfnamefont
  {S.}~\bibnamefont {Schmid}}, \ and\ \bibinfo {author} {\bibfnamefont
  {A.}~\bibnamefont {Boisen}},\ }\bibfield  {title} {\enquote {\bibinfo {title}
  {{Nanomechanical Infrared Spectroscopy with Vibrating Filters for
  Pharmaceutical Analysis}},}\ }\href {\doibase 10.1002/anie.201700052}
  {\bibfield  {journal} {\bibinfo  {journal} {Angewandte Chemie International
  Edition}\ }\textbf {\bibinfo {volume} {56}},\ \bibinfo {pages} {3901--3905}
  (\bibinfo {year} {2017})}\BibitemShut {NoStop}%
\bibitem [{\citenamefont {Schmid}\ \emph {et~al.}(2014)\citenamefont {Schmid},
  \citenamefont {Wu}, \citenamefont {Larsen}, \citenamefont {Rindzevicius},\
  and\ \citenamefont {Boisen}}]{Schmid2014d}%
  \BibitemOpen
  \bibfield  {author} {\bibinfo {author} {\bibfnamefont {S.}~\bibnamefont
  {Schmid}}, \bibinfo {author} {\bibfnamefont {K.}~\bibnamefont {Wu}}, \bibinfo
  {author} {\bibfnamefont {P.~E.}\ \bibnamefont {Larsen}}, \bibinfo {author}
  {\bibfnamefont {T.}~\bibnamefont {Rindzevicius}}, \ and\ \bibinfo {author}
  {\bibfnamefont {A.}~\bibnamefont {Boisen}},\ }\bibfield  {title} {\enquote
  {\bibinfo {title} {{Low-power photothermal probing of single plasmonic
  nanostructures with nanomechanical string resonators.}}}\ }\href {\doibase
  10.1021/nl4046679} {\bibfield  {journal} {\bibinfo  {journal} {Nano Letters}\
  }\textbf {\bibinfo {volume} {14}},\ \bibinfo {pages} {2318--2321} (\bibinfo
  {year} {2014})}\BibitemShut {NoStop}%
\bibitem [{\citenamefont {Larsen}\ \emph {et~al.}(2013)\citenamefont {Larsen},
  \citenamefont {Schmid}, \citenamefont {Villanueva},\ and\ \citenamefont
  {Boisen}}]{Larsen2013ACSNano}%
  \BibitemOpen
  \bibfield  {author} {\bibinfo {author} {\bibfnamefont {T.}~\bibnamefont
  {Larsen}}, \bibinfo {author} {\bibfnamefont {S.}~\bibnamefont {Schmid}},
  \bibinfo {author} {\bibfnamefont {L.~G.}\ \bibnamefont {Villanueva}}, \ and\
  \bibinfo {author} {\bibfnamefont {A.}~\bibnamefont {Boisen}},\ }\bibfield
  {title} {\enquote {\bibinfo {title} {{Photothermal analysis of individual
  nanoparticulate samples using micromechanical resonators.}}}\ }\href
  {\doibase 10.1021/nn402057f} {\bibfield  {journal} {\bibinfo  {journal} {ACS
  Nano}\ }\textbf {\bibinfo {volume} {7}},\ \bibinfo {pages} {6188--6193}
  (\bibinfo {year} {2013})}\BibitemShut {NoStop}%
\bibitem [{\citenamefont {Chien}\ \emph {et~al.}(2018)\citenamefont {Chien},
  \citenamefont {Brameshuber}, \citenamefont {Rossboth}, \citenamefont
  {Sch{\"u}tz},\ and\ \citenamefont {Schmid}}]{Chien2018}%
  \BibitemOpen
  \bibfield  {author} {\bibinfo {author} {\bibfnamefont {M.-H.}\ \bibnamefont
  {Chien}}, \bibinfo {author} {\bibfnamefont {M.}~\bibnamefont {Brameshuber}},
  \bibinfo {author} {\bibfnamefont {B.~K.}\ \bibnamefont {Rossboth}}, \bibinfo
  {author} {\bibfnamefont {G.~J.}\ \bibnamefont {Sch{\"u}tz}}, \ and\ \bibinfo
  {author} {\bibfnamefont {S.}~\bibnamefont {Schmid}},\ }\bibfield  {title}
  {\enquote {\bibinfo {title} {Single-molecule optical absorption imaging by
  nanomechanical photothermal sensing},}\ }\href {\doibase
  10.1073/pnas.1804174115} {\bibfield  {journal} {\bibinfo  {journal}
  {Proceedings of the National Academy of Sciences}\ }\textbf {\bibinfo
  {volume} {115}},\ \bibinfo {pages} {11150--11155} (\bibinfo {year} {2018})},\
  \Eprint
  {http://arxiv.org/abs/https://www.pnas.org/content/115/44/11150.full.pdf}
  {https://www.pnas.org/content/115/44/11150.full.pdf} \BibitemShut {NoStop}%
\bibitem [{\citenamefont {Piller}\ \emph {et~al.}(2019)\citenamefont {Piller},
  \citenamefont {Luhmann}, \citenamefont {Chien},\ and\ \citenamefont
  {Schmid}}]{Piller2019}%
  \BibitemOpen
  \bibfield  {author} {\bibinfo {author} {\bibfnamefont {M.}~\bibnamefont
  {Piller}}, \bibinfo {author} {\bibfnamefont {N.}~\bibnamefont {Luhmann}},
  \bibinfo {author} {\bibfnamefont {M.-H.}\ \bibnamefont {Chien}}, \ and\
  \bibinfo {author} {\bibfnamefont {S.}~\bibnamefont {Schmid}},\ }\bibfield
  {title} {\enquote {\bibinfo {title} {{Nanoelectromechanical infrared
  detector}},}\ }in\ \href {\doibase 10.1117/12.2528416} {\emph {\bibinfo
  {booktitle} {Optical Sensing, Imaging, and Photon Counting: From X-Rays to
  THz 2019}}},\ Vol.\ \bibinfo {volume} {1108802}\ (\bibinfo  {publisher}
  {SPIE},\ \bibinfo {year} {2019})\BibitemShut {NoStop}%
\bibitem [{\citenamefont {Zhang}\ \emph {et~al.}(2019)\citenamefont {Zhang},
  \citenamefont {Giroux}, \citenamefont {Nour},\ and\ \citenamefont
  {St-Gelais}}]{Zhang2019}%
  \BibitemOpen
  \bibfield  {author} {\bibinfo {author} {\bibfnamefont {C.}~\bibnamefont
  {Zhang}}, \bibinfo {author} {\bibfnamefont {M.}~\bibnamefont {Giroux}},
  \bibinfo {author} {\bibfnamefont {T.~A.}\ \bibnamefont {Nour}}, \ and\
  \bibinfo {author} {\bibfnamefont {R.}~\bibnamefont {St-Gelais}},\ }\bibfield
  {title} {\enquote {\bibinfo {title} {{Thermal radiation sensing using high
  mechanical Q-factor silicon nitride membranes}},}\ }\href {\doibase
  10.1109/SENSORS43011.2019.8956551} {\bibfield  {journal} {\bibinfo  {journal}
  {Proceedings of IEEE Sensors}\ }\textbf {\bibinfo {volume} {2019-Octob}}
  (\bibinfo {year} {2019}),\ 10.1109/SENSORS43011.2019.8956551}\BibitemShut
  {NoStop}%
\bibitem [{\citenamefont {Zhang}\ \emph {et~al.}(2013)\citenamefont {Zhang},
  \citenamefont {Myers}, \citenamefont {Sader},\ and\ \citenamefont
  {Roukes}}]{Zhang2013}%
  \BibitemOpen
  \bibfield  {author} {\bibinfo {author} {\bibfnamefont {X.~C.}\ \bibnamefont
  {Zhang}}, \bibinfo {author} {\bibfnamefont {E.~B.}\ \bibnamefont {Myers}},
  \bibinfo {author} {\bibfnamefont {J.~E.}\ \bibnamefont {Sader}}, \ and\
  \bibinfo {author} {\bibfnamefont {M.~L.}\ \bibnamefont {Roukes}},\ }\bibfield
   {title} {\enquote {\bibinfo {title} {{Nanomechanical Torsional Resonators
  for Frequency-Shift Infrared Thermal Sensing}},}\ }\href {\doibase
  10.1021/nl304687p} {\bibfield  {journal} {\bibinfo  {journal} {Nano Letters}\
  }\textbf {\bibinfo {volume} {13}},\ \bibinfo {pages} {1528--1534} (\bibinfo
  {year} {2013})}\BibitemShut {NoStop}%
\bibitem [{\citenamefont {Zhang}\ \emph {et~al.}(2020)\citenamefont {Zhang},
  \citenamefont {Giroux}, \citenamefont {Nour},\ and\ \citenamefont
  {St-Gelais}}]{Zhang2020}%
  \BibitemOpen
  \bibfield  {author} {\bibinfo {author} {\bibfnamefont {C.}~\bibnamefont
  {Zhang}}, \bibinfo {author} {\bibfnamefont {M.}~\bibnamefont {Giroux}},
  \bibinfo {author} {\bibfnamefont {T.~A.}\ \bibnamefont {Nour}}, \ and\
  \bibinfo {author} {\bibfnamefont {R.}~\bibnamefont {St-Gelais}},\ }\href@noop
  {} {\enquote {\bibinfo {title} {Radiative heat transfer in free-standing
  silicon nitride membranes},}\ } (\bibinfo {year} {2020}),\ \Eprint
  {http://arxiv.org/abs/2002.09017} {arXiv:2002.09017 [physics.app-ph]}
  \BibitemShut {NoStop}%
\bibitem [{\citenamefont {Larsen}\ \emph {et~al.}(2011)\citenamefont {Larsen},
  \citenamefont {Schmid}, \citenamefont {Gr{\"{o}}nberg}, \citenamefont
  {Niskanen}, \citenamefont {Hassel}, \citenamefont {Dohn},\ and\ \citenamefont
  {Boisen}}]{Larsen2011}%
  \BibitemOpen
  \bibfield  {author} {\bibinfo {author} {\bibfnamefont {T.}~\bibnamefont
  {Larsen}}, \bibinfo {author} {\bibfnamefont {S.}~\bibnamefont {Schmid}},
  \bibinfo {author} {\bibfnamefont {L.}~\bibnamefont {Gr{\"{o}}nberg}},
  \bibinfo {author} {\bibfnamefont {A.~O.}\ \bibnamefont {Niskanen}}, \bibinfo
  {author} {\bibfnamefont {J.}~\bibnamefont {Hassel}}, \bibinfo {author}
  {\bibfnamefont {S.}~\bibnamefont {Dohn}}, \ and\ \bibinfo {author}
  {\bibfnamefont {A.}~\bibnamefont {Boisen}},\ }\bibfield  {title} {\enquote
  {\bibinfo {title} {{Ultrasensitive string-based temperature sensors}},}\
  }\href {\doibase 10.1063/1.3567012} {\bibfield  {journal} {\bibinfo
  {journal} {Applied Physics Letters}\ }\textbf {\bibinfo {volume} {98}},\
  \bibinfo {pages} {121901} (\bibinfo {year} {2011})}\BibitemShut {NoStop}%
\bibitem [{\citenamefont {Bergman}\ \emph {et~al.}(2017)\citenamefont
  {Bergman}, \citenamefont {Lavine}, \citenamefont {Incropera},\ and\
  \citenamefont {DeWitt}}]{Bergman2017}%
  \BibitemOpen
  \bibfield  {author} {\bibinfo {author} {\bibfnamefont {T.~L.}\ \bibnamefont
  {Bergman}}, \bibinfo {author} {\bibfnamefont {A.~S.}\ \bibnamefont {Lavine}},
  \bibinfo {author} {\bibfnamefont {F.~P.}\ \bibnamefont {Incropera}}, \ and\
  \bibinfo {author} {\bibfnamefont {D.~P.}\ \bibnamefont {DeWitt}},\
  }\href@noop {} {\emph {\bibinfo {title} {{Fundamentals of Heat and Mass
  Transfer}}}}\ (\bibinfo  {publisher} {Wiley},\ \bibinfo {year}
  {2017})\BibitemShut {NoStop}%
\bibitem [{\citenamefont {Toivola}\ \emph {et~al.}(2003)\citenamefont
  {Toivola}, \citenamefont {Thurn}, \citenamefont {Cook}, \citenamefont
  {Cibuzar},\ and\ \citenamefont {Roberts}}]{Toivola2003}%
  \BibitemOpen
  \bibfield  {author} {\bibinfo {author} {\bibfnamefont {Y.}~\bibnamefont
  {Toivola}}, \bibinfo {author} {\bibfnamefont {J.}~\bibnamefont {Thurn}},
  \bibinfo {author} {\bibfnamefont {R.~F.}\ \bibnamefont {Cook}}, \bibinfo
  {author} {\bibfnamefont {G.}~\bibnamefont {Cibuzar}}, \ and\ \bibinfo
  {author} {\bibfnamefont {K.}~\bibnamefont {Roberts}},\ }\bibfield  {title}
  {\enquote {\bibinfo {title} {{Influence of deposition conditions on
  mechanical properties of low-pressure chemical vapor deposited low-stress
  silicon nitride films}},}\ }\href {\doibase 10.1063/1.1622776} {\bibfield
  {journal} {\bibinfo  {journal} {Journal of Applied Physics}\ }\textbf
  {\bibinfo {volume} {94}},\ \bibinfo {pages} {6915--6922} (\bibinfo {year}
  {2003})}\BibitemShut {NoStop}%
\bibitem [{\citenamefont {Ftouni}\ \emph {et~al.}(2015)\citenamefont {Ftouni},
  \citenamefont {Blanc}, \citenamefont {Tainoff}, \citenamefont {Fefferman},
  \citenamefont {Defoort}, \citenamefont {Lulla}, \citenamefont {Richard},
  \citenamefont {Collin},\ and\ \citenamefont {Bourgeois}}]{Ftouni2015}%
  \BibitemOpen
  \bibfield  {author} {\bibinfo {author} {\bibfnamefont {H.}~\bibnamefont
  {Ftouni}}, \bibinfo {author} {\bibfnamefont {C.}~\bibnamefont {Blanc}},
  \bibinfo {author} {\bibfnamefont {D.}~\bibnamefont {Tainoff}}, \bibinfo
  {author} {\bibfnamefont {A.~D.}\ \bibnamefont {Fefferman}}, \bibinfo {author}
  {\bibfnamefont {M.}~\bibnamefont {Defoort}}, \bibinfo {author} {\bibfnamefont
  {K.~J.}\ \bibnamefont {Lulla}}, \bibinfo {author} {\bibfnamefont
  {J.}~\bibnamefont {Richard}}, \bibinfo {author} {\bibfnamefont
  {E.}~\bibnamefont {Collin}}, \ and\ \bibinfo {author} {\bibfnamefont
  {O.}~\bibnamefont {Bourgeois}},\ }\bibfield  {title} {\enquote {\bibinfo
  {title} {{Thermal conductivity of silicon nitride membranes is not sensitive
  to stress}},}\ }\href {\doibase 10.1103/PhysRevB.92.125439} {\bibfield
  {journal} {\bibinfo  {journal} {Physical Review B - Condensed Matter and
  Materials Physics}\ }\textbf {\bibinfo {volume} {92}},\ \bibinfo {pages}
  {1--7} (\bibinfo {year} {2015})}\BibitemShut {NoStop}%
\bibitem [{\citenamefont {Feng}, \citenamefont {Li},\ and\ \citenamefont
  {Zhang}(2009)}]{Feng2009}%
  \BibitemOpen
  \bibfield  {author} {\bibinfo {author} {\bibfnamefont {B.}~\bibnamefont
  {Feng}}, \bibinfo {author} {\bibfnamefont {Z.}~\bibnamefont {Li}}, \ and\
  \bibinfo {author} {\bibfnamefont {X.}~\bibnamefont {Zhang}},\ }\bibfield
  {title} {\enquote {\bibinfo {title} {{Prediction of size effect on thermal
  conductivity of nanoscale metallic films}},}\ }\href {\doibase
  10.1016/j.tsf.2008.10.116} {\bibfield  {journal} {\bibinfo  {journal} {Thin
  Solid Films}\ }\textbf {\bibinfo {volume} {517}},\ \bibinfo {pages}
  {2803--2807} (\bibinfo {year} {2009})}\BibitemShut {NoStop}%
\bibitem [{\citenamefont {Lugo}\ and\ \citenamefont {Oliva}(2016)}]{Lugo2016}%
  \BibitemOpen
  \bibfield  {author} {\bibinfo {author} {\bibfnamefont {J.~M.}\ \bibnamefont
  {Lugo}}\ and\ \bibinfo {author} {\bibfnamefont {A.~I.}\ \bibnamefont
  {Oliva}},\ }\bibfield  {title} {\enquote {\bibinfo {title} {{Thermal
  Properties of Metallic Films at Room Conditions by the Heating Slope}},}\
  }\href {\doibase 10.2514/1.T4605} {\bibfield  {journal} {\bibinfo  {journal}
  {Journal of Thermophysics and Heat Transfer}\ }\textbf {\bibinfo {volume}
  {30}},\ \bibinfo {pages} {452--460} (\bibinfo {year} {2016})}\BibitemShut
  {NoStop}%
\bibitem [{\citenamefont {Kruse}, \citenamefont {McGlauchlin},\ and\
  \citenamefont {McQuistan}(1962)}]{kruse1962elements}%
  \BibitemOpen
  \bibfield  {author} {\bibinfo {author} {\bibfnamefont {P.}~\bibnamefont
  {Kruse}}, \bibinfo {author} {\bibfnamefont {L.}~\bibnamefont {McGlauchlin}},
  \ and\ \bibinfo {author} {\bibfnamefont {R.}~\bibnamefont {McQuistan}},\
  }\href {https://books.google.at/books?id=PCVRAAAAMAAJ} {\emph {\bibinfo
  {title} {Elements of infrared technology: generation, transmission, and
  detection}}}\ (\bibinfo  {publisher} {Wiley},\ \bibinfo {year}
  {1962})\BibitemShut {NoStop}%
\end{thebibliography}%
\clearpage
\onecolumngrid
\subsection*{SUPPLEMENTARY INFORMATION}
\subsection{Calculation of emissivity}
In order to obtain the emissivity of our SiN drums we employed Kirchoff's law for a non ideal radiator, where the emissivity corresponds to the absorption, $\varepsilon = \alpha$.\cite{kruse1962elements} 

\begin{figure} [htbp]
\includegraphics{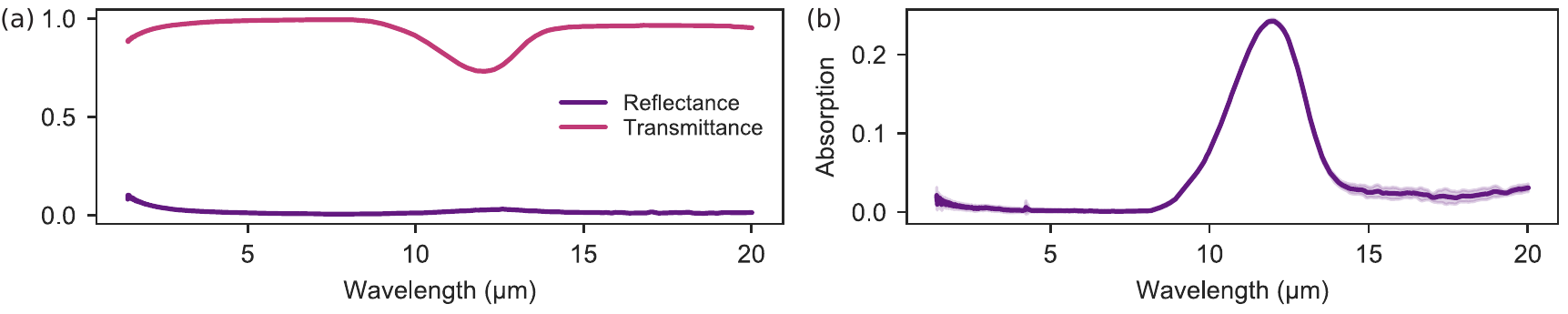} 
\caption{(a) Measured transmittance $\tau$ and reflectance $\rho$ spectra for a \SI{50}{\nm} thick SiN drum. (b) Absorption spectrum of the SiN layer.}
\label{fig:sfig1}
\end{figure}

Therefore, we needed to measure the optical spectra via Fourier transformed IR spectroscopy (Bruker Tensor 27) with a specified transmittance and reflectance unit (Bruker A510/Q-T).
From the measured transmittance and reflectance spectra, as shown in Fig.~\ref{fig:sfig1}(a), we calculated the absorption spectrum by $\alpha = 1 - \rho - \tau$, which is shown in Fig.~\ref{fig:sfig1}(b), where $\rho$ is the reflectance and $\tau$ the transmittance.

To obtain the emissivity at $T = 300K$ we first calculated the total emissive power of a blackbody $E_b$ as~\cite{Bergman2017}

\begin{equation}
E_{b}=\int_{0}^{\infty} \frac{2 \pi h c_{0}^{2}}{\lambda^{5}\left[\exp \left(\left(h c_{o} / k\right) / \lambda T\right)-1\right]} d \lambda,
\end{equation}

where $h$ the Planck constant, $k$ the Boltzmann constant, $c_0$ the speed of light and $\lambda$ the wavelength. 
Then the spectral emissive power $E_{\lambda,b}(\lambda,T)$ is calculated as following~\cite{Bergman2017}

\begin{equation}
E_{\lambda, b}(\lambda, T)=\frac{2 \pi h c_{0}^{2}}{\lambda^{5}\left[\exp \left(\left(h c_{o} / k\right) / \lambda T\right)-1\right]}.
\end{equation}

Considering the absorption $\alpha$ as $\varepsilon_\lambda(\lambda,T)$ we can calculate the effective emissivity by

\begin{equation}
\varepsilon(T)=\frac{\int_{0}^{\infty} \varepsilon_{\lambda}(\lambda, T) E_{\lambda,b}(\lambda, T) d \lambda}{E_{b}(T)}
\end{equation}

\begin{figure}[htbp]
\includegraphics{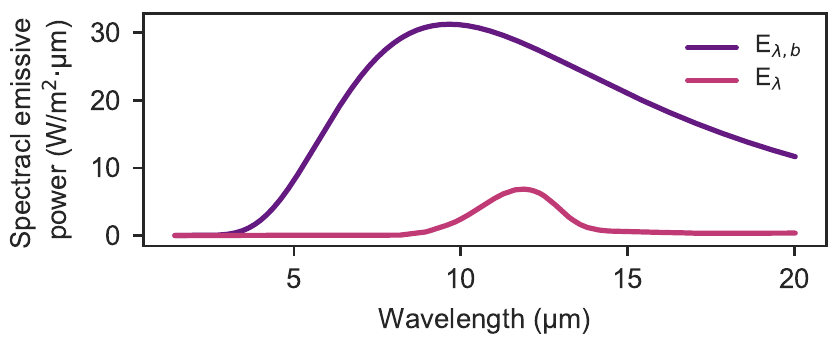} 
\caption{Comparison of the SiN surface emission $E_\lambda = \varepsilon_{\lambda}(\lambda, T) E_{\lambda, b}(\lambda, T)$ and of a blackbody $E_\lambda,b$.}
\label{fig:sfig2}
\end{figure}
In our case we integrated over the measured spectrum from \SI{1.4}{\um} to \SI{20}{\um} and obtained a emissivity $\varepsilon \approx 0.05$.
Fig.~\ref{fig:sfig2} shows the spectra for the measured wavelength range.

\newpage
\subsection{Response time measurements}
For response time measurements the frequency of the modulated heating laser is constantly increased while recording the frequency via the phase-locked loop. 
At low values of the modulation frequency, the drum resonator can follow this modulation. 
However, if the modulation frequency of the heating increases, the resonator is, due to its material properties and thermal transfer capabilities, not able to follow. 
The amplitude of the recorded time signal will decrease as a consequence. 
\begin{figure}[htb]
\includegraphics[width=0.35\textwidth]{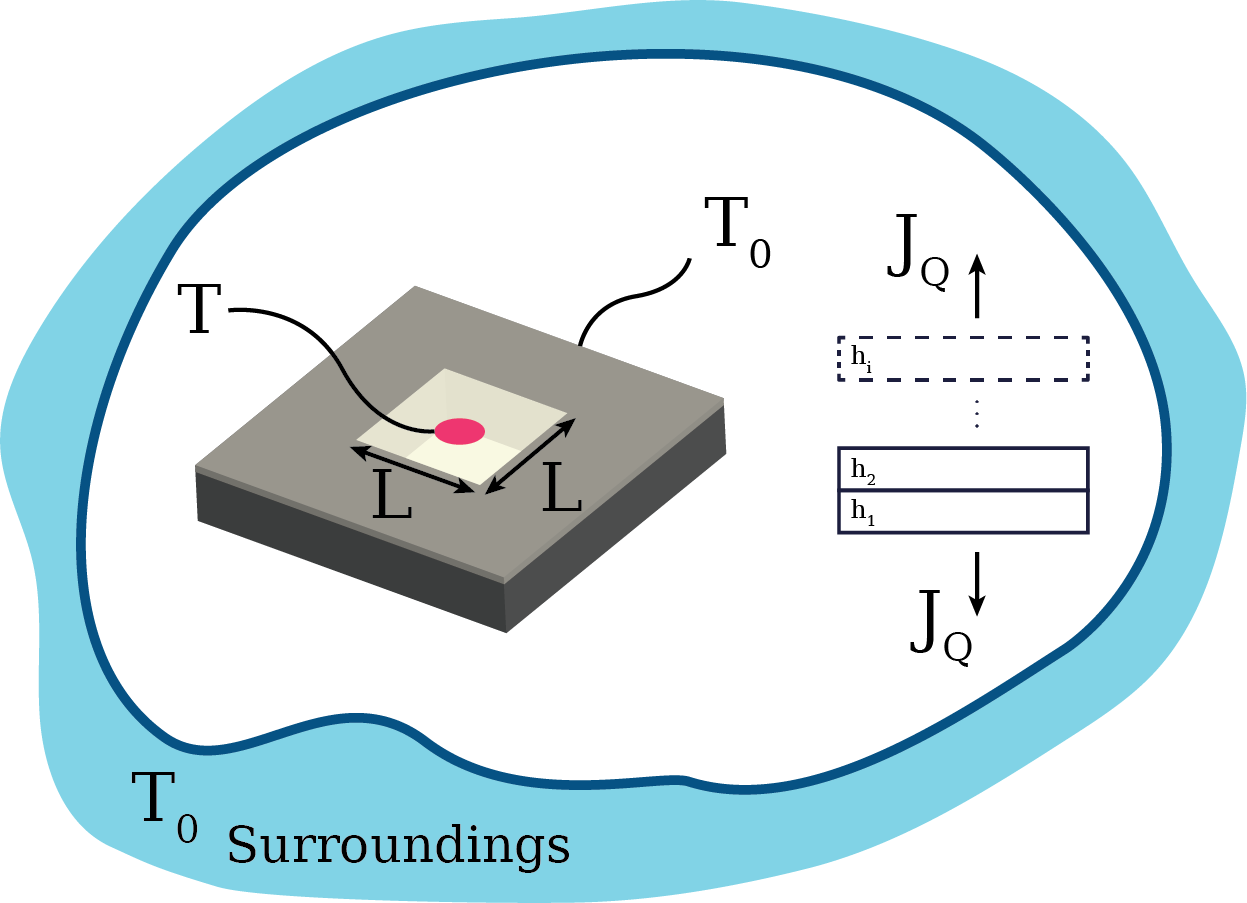} 
\caption{\label{fig:sfig3} Schematic depiction of quantities for the mathematical derivation.}
\end{figure}

To compare the results from measurements we derived an analytical model, assuming a setting similar to the schematic in Fig.~\ref{fig:sfig3}. The drum size is given by its lateral dimension $L$ and we assume a constant surrounding temperature $T_0$ that also acts as a boundary condition for the temperature of the silicon frame supporting the silicon nitride and additional layers. 

Each layer by index $i \in [1,2,...,n]$ has a certain height given by $h_i$, a material specific thermal conductivity $\kappa_i$ and a heat capacity per volume $c_i$.
 The temperature $T$ in the center of the drum resonator originates from absorbed photons with an absorbed power $P$.
Depending on the pressure $p$ there might be a heat loss to the ambient gas which is given by some heat flux density $J_Q$. 
We denote the origin from the heat flux density by its subscript $i$, e.g. $J_{Q1}$ for the bottom side and $J_{Qn}$ for the top side of the drum.

The heat continuity equation is given by

\begin{equation}
\label{eq:one}
    c_{i} \frac{\partial T}{\partial t}=-\nabla \cdot \mathbf{J}_{Q i}+q(r, t)=\nabla \cdot\left(\kappa_{i} \nabla T\right)+q(r, t) \simeq \kappa_{i} \nabla_{\perp}^{2} T+\frac{\partial}{\partial z}\left(\kappa_{i} \frac{\partial T}{\partial z}\right)+q(r, t)
\end{equation}
where $q(r, t)$ is the dissipated heat rate per unit volume, $\mathbf{J}_{Qi}=-\kappa_{i} \nabla T$ the heat current density.
The temperature in the vertical direction $z$ can be assumed to be constant due to the great lateral dimension compared to the thickness. 
Integrating equation~(\ref{eq:one}) over the thickness leads to,

\begin{equation}
    \left(\sum_{i} h_{i} c_{i}\right) \frac{\partial T}{\partial t}=\left(\sum_{i} h_{i} \kappa_{i}\right) \nabla_{\perp}^{2} T-J_{Qn}-J_{Q1}+Q(x, y, t).
\end{equation}

Within a certain range of ambient pressures the heat flux $J_{Qi}$ can be expected to vary linearly with the drum temperature when each molecule is expected to remove energy by its impinging molecular flux density $\frac{1}{4} n v$, where $n=p /\left(k_{B} T_{0}\right)$ is the molecular density and $v$ the mean molecular speed. The removed energy is given by $\frac{f}{2} k_{B}(T-T_{0})$, where $f$ is the number of degrees of freedom for the molecules (e.g. $f = 6$).
The heat flux is assumed to be equal on the top and bottom surface given by

\begin{equation}\label{eq:3}
    J_{Qn}=J_{Q \text {1}}=\frac{1}{4} n v \frac{f}{2} k_{B}\left(T-T_{0}\right)=\frac{p v f}{8 T_{0}}\left(T-T_{0}\right)=\frac{1}{2} H\left(T-T_{0}\right).
\end{equation}

With equation~(\ref{eq:3}) the equation~(\ref{eq:one}) governing the temperature becomes
\begin{equation}
\left(\sum_{i} h_{i} c_{i}\right) \frac{\partial T}{\partial t}=\left(\sum_{i} h_{i} \kappa_{i}\right) \nabla_{\perp}^{2} T-H\left(T-T_{0}\right)+Q(x, y, t).
\end{equation}

For the sake of simplicity, the temperature $T_0$ is set to zero. The temperatures $T$ can now be seen as the deviating temperatures.
It follows that

\begin{align}
    C \frac{\partial T}{\partial t} &= K \nabla_{\perp}^{2} T-H T+Q(x, y, t) \\
    C &=\sum_{i} h_{i} c_{i}\quad \text{ and } \quad K=\sum_{i} h_{i} \kappa_{i}
\end{align}

\begin{align}
    \frac{1}{D} \frac{\partial T}{\partial t} &=\nabla_{\perp}^{2} T-\frac{H}{K} T+\frac{Q(x, y, t)}{K} \\
    D &=\frac{\sum_{i} h_{i} \kappa_{i}}{\sum_{i} h_{i} c_{i}}.
\end{align}

For symmetric solutions in $x$ and $y$ we can assume product solutions in the form of 
\begin{equation}
 \varphi_{n}=A_{n} \cos \left(k_{n} x\right) \cos \left(k_{n} y\right) \exp \left(-\frac{t}{\tau_{n}}\right)
\end{equation}

with $\cos \left(k_{n} L / 2\right)=0 \Rightarrow k_{n}= \pi(2 n+1)/L$.
The homogeneous solution of the partial differential equation is then
\begin{equation}
\begin{aligned}
-\frac{1}{D \tau_{n}} &=-2 k_{n}^{2}-\frac{H}{K} \Rightarrow \frac{1}{\tau_{n}}=D\left(2 k_{n}^{2}+\frac{H}{K}\right) \\
\tau_{n} &=\frac{1}{D\left(2 k_{n}^{2}+\frac{H}{K}\right)}.
\end{aligned}
\end{equation}

Finally we are interested in the first mode ($n = 0$), since for higher modes the decay is faster and obtain

\begin{equation}\label{eq:18}
\begin{aligned}
\frac{1}{\tau_{0}} &=D\left(2 k_{0}^{2}+\frac{H}{K}\right)=D\left(\frac{2 \pi^{2}}{L^{2}}+\frac{H}{K}\right) \\
&=\frac{2 \pi^{2}}{L^{2}} \frac{\sum_{i} h_{i} \kappa_{i}}{\sum_{i} h_{i} c_{i}}+\frac{H}{\sum_{i} h_{i} c_{i}}.
\end{aligned}
\end{equation}

Dependent on the quality of the vacuum, (18) can be reduced by assuming $H\simeq0$, describing a thermal conductivity dominated heat transfer. However, to include thermal radiation, the heat flux density from the top and bottom surface needs to be considered as 
\begin{equation}
J_{Qn}=J_{Q1}=\epsilon\sigma_{\mathrm{SB}}\left(  T^{4}-T_{0}^{4}\right).
\end{equation} 
For small deviations from the ambient temperature, the nonlinear temperature dependency can be replaced by a first-order Taylor expansion with reasonable accuracy given by 
\begin{equation}
J_{Qn}=J_{Q1}\simeq4\epsilon\sigma_{\mathrm{SB}}T_{0}^{3}\left(
T-T_{0}\right)  =\frac{1}{2}H_{\mathrm{SB}}\left(  T-T_{0}\right).
\end{equation}
The thermal radiation can then be accounted for by replacing $H$ in equation (18) with $H_{\mathrm{SB}}=$ $8\epsilon\sigma_{\mathrm{SB}}T_{0}^{3}$, i.e.,
\begin{equation}
\frac{1}{\tau_{0}}=\frac{2\pi^{2}}{L^{2}}\frac{\sum_{i}h_{i}\kappa_{i}}%
{\sum_{i}h_{i}c_{i}}+\frac{8\epsilon\sigma_{\mathrm{SB}}T_{0}^{3}}{\sum
_{i}h_{i}c_{i}}.
\end{equation}

\end{document}